\documentclass[preprint, 12pt, authoryear]{elsarticle}
\pdfoutput=1
\usepackage{etoolbox}
\renewcommand*{\today}{August 29, 2018}
\patchcmd{\MaketitleBox}{\footnotesize\itshape\elsaddress\par\vskip36pt}{\footnotesize\itshape\elsaddress\par\parbox[b][36pt]{\linewidth}{\vfill\hfill\textnormal{\today}\hfill\null\vfill}}{}{}%
\patchcmd{\pprintMaketitle}{\footnotesize\itshape\elsaddress\par\vskip36pt}{\footnotesize\itshape\elsaddress\par\parbox[b][36pt]{\linewidth}{\vfill\hfill\textnormal{\today}\hfill\null\vfill}}{}{}%

\usepackage{eurosym} 
\usepackage{gensymb}
\usepackage{makecell}
\usepackage{subcaption}
\usepackage{url}
\usepackage{breakurl}

\usepackage[version=4]{mhchem} 
\usepackage{amsmath} 
\usepackage{amsfonts}
\usepackage{amssymb}
\usepackage{graphicx} 

\makeatletter
\def\ps@pprintTitle{%
 \let\@oddhead\@empty
 \let\@evenhead\@empty
 \def\@oddfoot{}%
 \let\@evenfoot\@oddfoot}
\makeatother

\begin{document}
\begin{frontmatter}
\title{District heating systems under high \ce{CO2} emission prices: the role of the pass-through from emission cost to electricity prices}

\author[1,2]{Sebastian Wehrle\corref{cor1}}
\ead{sebastian.wehrle@boku.ac.at}

\author[1]{Johannes Schmidt}
\ead{johannes.schmidt@boku.ac.at}

\cortext[cor1]{Corresponding author}

\address[1]{Institute for Sustainable Economic Development, University of Natural Resources and Life Sciences, Feistmantelstrasse 4, 1180 Vienna, Austria}

\address[2]{Wiener Netze GmbH, Erdbergstrasse 236, 1110 Vienna, Austria}

\begin{abstract}
Low \ce{CO2} prices have prompted discussion about political measures aimed at increasing the cost of carbon dioxide emissions.
These costs affect, inter alia, integrated district heating system operators (DHSO), often owned by municipalities with some political influence, that use a variety of (\ce{CO2} emission intense) heat generation technologies.
We examine whether DHSOs have an incentive to support measures that increase \ce{CO2} emission prices in the short term.
Therefore, we (i) develop a simplified analytical framework to analyse optimal decisions of a district heating operator, and (ii) investigate the market-wide effects of increasing emission prices, in particular the pass-through from emission costs to electricity prices.
Using a numerical model of the common Austrian and German power system, we estimate a pass-through from \ce{CO2} emission prices to power prices between 0.69 and 0.53 as of 2017, depending on the absolute emission price level.
We find the \ce{CO2} emission cost pass-through to be sufficiently high so that low-emission district heating systems operating at least moderately efficient generation units benefit from rising \ce{CO2} emission prices in the short term.
\end{abstract}

\end{frontmatter}

\section{Introduction} \label{intro}
With the ratification of the Paris Agreement, the European Union (EU) committed itself to pursue climate policies to limit global warming to 'well below 2\degree C'~\citep{UNFCCC2015}, which implies the need for a drastic reduction in global greenhouse gas (GHG) emissions.
In Europe, the EU introduced the Emissions Trading System (ETS) as its flagship instrument to reduce GHG emissions.
Under the EU ETS, a cap is set on the maximum amount of GHG that can be emitted.
A corresponding amount of emission allowances (EUA) is allocated and can be traded on exchanges ('cap-and-trade').
In 2016, approximately 40\% of total greenhouse gas emissions in 31 countries were covered by the EU ETS~\citep{EEA2017}.
In particular, the EU ETS covers carbon dioxide (\ce{CO2}) emissions from power generators, refineries, a range of energy-intensive industries, and commercial airlines.

Over the recent years, EUA prices declined from an average of $23.19$ Euro per metric ton of \ce{CO2} equivalent (\euro{}/t\ce{CO2}) in 2008 to an average of $5.89$ \euro{}/t\ce{CO2} in 2017 (eex, 2018).
In response to low emission prices, the EU decided to rein in emission allowance supply through 'backloading' and the 'market stability reserve'.
While both measures are directed at increasing prices, they are expected to be effective only after 2019.

Several EU member states aim at reducing GHG emissions significantly.
Germany, for example, seeks to cut GHG emissions by 80\% to 95\% below 1990 levels by 2050~\citep{BMWi2010}.
However, the incentives for emission reduction provided by current \ce{CO2} emission prices are considered insufficient to reach ambitious long-term GHG reduction goals, not only by the German government~\citep{BMU2014}.
In consequence, additional measures to support further GHG emission reductions are under consideration.

In the following, we examine whether district heating and cooling system operators (DHSO), often municipal utilities, have an incentive to support measures that increase \ce{CO2} emission prices in the short term.

\citet{Jouvet2013} show that power generators with low or no carbon dioxide emissions can benefit from rents that are created by the pass-through of emission costs to electricity prices.
To the best of our knowledge, so far there is no literature explicitly considering the effects of (changes in) emission costs on district heating systems.
Therefore, we (i) develop a simple, stylized analytical framework to analyse decisions of a district heating operator, and (ii) investigate the market-wide effects of increasing emission costs, in particular the pass-through of \ce{CO2} emission prices to electricity prices with the power system model \emph{MEDEA\_lin}.
While we rely on a power system model to quantify the cost pass-through at varying levels of \ce{CO2} emission prices, the corresponding literature is dominated by econometric analysis~\citep{Sijm2006, Zachmann2008, Hintermann2016, Fabra2014} of pass-through levels conditional on realized historical emission prices.

\begin{table}
\caption{Summary of emission cost pass-through reported in the literature}
\smallskip
\centering
\begin{tabular}{l c c c c c}
\hline
Author(s) & \makecell{Estimated \\ pass-\\ through} & \makecell{Std.\\ Dev} & \makecell{Avg. \\ Emission \\ Price} & \makecell{Time\\ Period} & Country\\ \hline \hline
\makecell[l]{Fabra and\\ Reguant (2014)} & \makecell{$0.862$ /\\ $0.835$} & \makecell{$0.181$ /\\ $0.173$} & $19.12$ & \makecell{2004-\\2006} & Spain\\ \hline
\citet{Hintermann2016} & $0.962$ & $0.0753$ & $9.79$ & \makecell{2010-\\ 2013} & Germany\\ \hline
\citet{Sijm2006} & $1.17$ / $0.6$ & NA & $18.25$ & 2005 & Germany \\ \hline \hline
\end{tabular}
\label{pttab}
\end{table}

\citet{Fabra2014} conduct an econometric analysis of the pass-through from \ce{CO2} prices to power prices.
A rich data set available for the Spanish power market allows estimating the impact of changes in hourly marginal cost on emission prices.
Depending on the model specification, the authors find an average pass-through rate between $0.835$ and $0.862$ over the course of January~2004 to February~2006.
According to \citet{Fabra2014}, the measured pass-through is explained by (i) weak incentives for markup adjustment, which is in turn explained by the high correlation of cost shocks among firms and by the limited demand elasticity, and (ii) the absence of relevant price rigidities.

\citet{Sijm2006} analyse potential windfall profits that accrue to power companies in the wake of the introduction of the EU ETS. The authors cautiously provide empirical estimates for the German electricity market in 2005, warning, however, of methodological difficulties related to their estimation strategy.
Pointing out potential underestimation, pass-through is evaluated within the range of $0.6$ to $1.17$. At this time, prices for EU emission allowances averaged $18.25$ \euro{}/t\ce{CO2}.

\citet{Hintermann2016} argues that econometric analyses based on price or price spread regressions produce biased pass-through estimates, amongst others due to the merit order being correlated with input prices.
To improve on price regressions, Hintermann constructs estimates of hourly marginal cost from detailed power sector data.
Using this dataset he finds pass-through rates between $0.81$ and $1.11$ for the German market from January~2010 through December~2013.

Apart from potential endogeneity issues, regression analysis is necessarily based on historical observations.
Extrapolation from historical data, which does not include (very) high emission prices, may lead to considerable bias, in particular in non-linear systems such as power systems.
Our modelling effort therefore complements econometric analysis and goes beyond the approach in  \citet{Hintermann2016}, as we take into account full (inter-temporal) optimization of the power system, instead of relying on hourly models of the merit order only.

\section{A stylized model of district heating operations}
\subsection{Cost minimization in district heating systems}
District heat is typically supplied by a broad portfolio of heat generation technologies.
Some of these technologies, for example heat boilers, are generating heat only, while others, such as combined heat and power (CHP) plants, couple the generation of heat and electricity.
A cost-minimizing operator will dispatch heat generation units according to their marginal cost, with the lowest cost unit being dispatched first.
For co-generation units, marginal cost is affected by the prevailing prices for electricity, fuels and \ce{CO2} emissions.
To analyse the effect of emission costs on the decisions of district heating suppliers, a simplified portfolio consisting of co-generation units $C$ and natural gas boilers $B$ is considered.
The cost function $K$ of the CHP operator is then given by
\begin{align}\label{1}
K = p_{f} q_{f}^{B} + p_{e} em_{f} q_{f}^{B} + p_{f} q_{f}^{C} + p_{e} em_{f} q_{f}^{C} - p_{el} \eta_{el} q_{f}^{C}
\end{align}
where $p$ denotes prices of fuel $f$ (e.g.~coal or natural gas), $e$ emissions of \ce{CO2}, $el$ stands for electricity, and $q$ denotes quantities of fuel used in boiler $B$ or CHP $C$.
Plant efficiency (measured as MWh of output per MWh of input) is denoted by $\eta$ and $em$ is the fuel emission factor in tons of \ce{CO2} emitted per MWh of fuel used.
At each point in time, heat generation from the portfolio has to match heat demand from the system, such that $q_{f}^{B} \eta_{th}^{B} + q_{f}^{C} \eta_{th}^{C} = d_{th}$.
The corresponding cost-minimization problem can be represented by the Langrangian
\begin{align}\label{2}
\begin{split}
\mathcal{L}(q_{f}^{B},q_{f}^{C},\lambda)= p_{f} q_{f}^{B} + p_{e} em_{f} q_{f}^{B} + p_{f} q_{f}^{C} + p_{e} em_{f} q_{f}^{C} - p_{el} \eta_{el} q_{f}^{C} \\ - \lambda ( q_{f}^{B} \eta_{th}^{B} + q_{f}^{C} \eta_{th}^{C} - d_{th})
\end{split}
\end{align}
The first-order conditions of the problem in (\ref{2}) allow to express shadow prices (cost $k$) of heat generated by a heat boiler and a CHP-plant, respectively.
\begin{align}\label{3_4}
k_{th}^{B} = \frac{1}{\eta_{th}^{B}} (p_{f} + p_{e} em_{f}) \\
k_{th}^{C} = \frac{1}{\eta_{th}^{C}} (p_{f} + p_{e} em_{f} - \eta_{el}^{C} p_{el})
\end{align}
As long as sufficient capacities are available (e.g.~in summer, when space heating demand is low), a cost minimizing operator will dispatch the lowest cost unit first.
CHP units will have an absolute cost advantage over boilers $(k_{th}^{C} \leq k_{th}^{B})$, if the 'mark-up' of the electricity price over fuel and \ce{CO2} emission cost is sufficiently high.
\begin{align}\label{5}
\frac{p_{el}}{(p_{f} + p_{e} em_{f})} \geq \frac{\eta_{th}^{B} - \eta_{th}^{C}}{\eta_{th}^{B} \eta_{el}^{C}}
\end{align}
Boilers will be dispatched when electricity prices are low (below the threshold in (\ref{5})) or when CHP units are capacity constrained (which is typically the case in winter when space heating demand peaks).

\subsection{Effect of rising \ce{CO2} prices on district heating systems}
Upon the introduction of measures to increase the price $p_{e}$ the \ce{CO2} emission cost of CHP units and boilers will increase.
Moreover, any measure that induces a general increase in \ce{CO2} emission prices will also affect generation costs of \ce{CO2} emitting power generators, who will pass some of the cost increase on to electricity prices $p_{el}$.
Thus, the electricity price depends, amongst others, on the emission price, i.e. $p_{el}(\cdot, p_{e})$.
Rising electricity prices generate additional revenues for CHP units from electricity sales.
If these additional revenues outweigh the additional generation cost, district heating companies benefit from rising \ce{CO2} prices.
In other words, the total cost of district heat generation $k^{tot} = k_{th}^{C} q_{th}^{C} + k_{th}^{B} q_{th}^{B}$ need to decline in emission prices.
This will be the case if
\begin{align}\label{6}
\frac{\partial k^{tot}}{\partial p_{e}} = \frac{\partial k_{th}^{C}}{\partial p_{e}} q_{th}^{C} + k_{th}^{C}
\frac{\partial q_{th}^C}{\partial p_e} + \frac{\partial k_{th}^{B}}{\partial p_{e}} q_{th}^{B} + k_{th}^{B}
\frac{\partial q_{th}^{B}}{\partial p_{e}} < 0
\end{align}
To simplify this expression, we make use of the fact that (exogenous) heat demand $d_{th}$ must be supplied entirely from the district heating system, such that any reduction in heat generation from one source must lead to a corresponding increase in heat generation of another source, i.e.
$q_{th}^{C} + q_{th}^{B} = d_{th} \Rightarrow \partial q^{B} /\partial p_{e} = - \partial q^{C} / \partial p_{e}$.
Based on this, we can rewrite the 'declining total cost-condition' in (\ref{6}) as
\begin{align}\label{7}
\frac{\partial k_{th}^{C}}{\partial p_{e}} q^{C} + \frac{\partial k_{th}^{B}}{\partial p_{e}} q^{B} < (k_{th}^{B} - k_{th}^{C})\frac{\partial q^{C}}{\partial p_{e}}
\end{align}
The inequality in (\ref{7}) relates produced quantities from boilers and CHPs to these units' heat cost.
Using the identity $q^{B} = q^{T} - q^{C}$, with $q^{T}$ being total heat production from emission intense generators, equation (\ref{7}) can be rearranged to yield the lower threshold on the pass-through from emission cost to electricity prices for which a district heating system with a given share of CHP heat generation and given electrical efficiency reduces its total cost.

In general, total costs of heat generation are declining in emission prices, if
\begin{align}\label{8}
\frac{\partial p_{el}}{\partial p_e} > \underbrace{\left( \frac{q^T - q^C}{q^C} \frac{\eta_{th}^{C}}{\eta_{th}^{B} \eta_{el}^{C}} + \frac{\eta_{th}^{B}}{\eta_{th}^{B} \eta_{el}^{C}} \right) em_{f}}_\text{A} \underbrace{- \frac{(k_{th}^{B} - k_{th}^{C})}{p_{el}} \varepsilon_{s^{C}}^{p_{el}} \frac{\eta_{th}^{C}}{\eta_{el}^{C}} \frac{\partial p_{el}}{\partial p_{e}}}_\text{B}
\end{align}
where $\varepsilon_{s^{C}}^{p_{el}} = (\partial q^{C} / \partial p_{el}) (p_{el} / q^{C}) \geq 0$ is the electricity price elasticity of heat supply from CHP units.
\begin{figure}[t]
\centering
\includegraphics[scale=0.35]{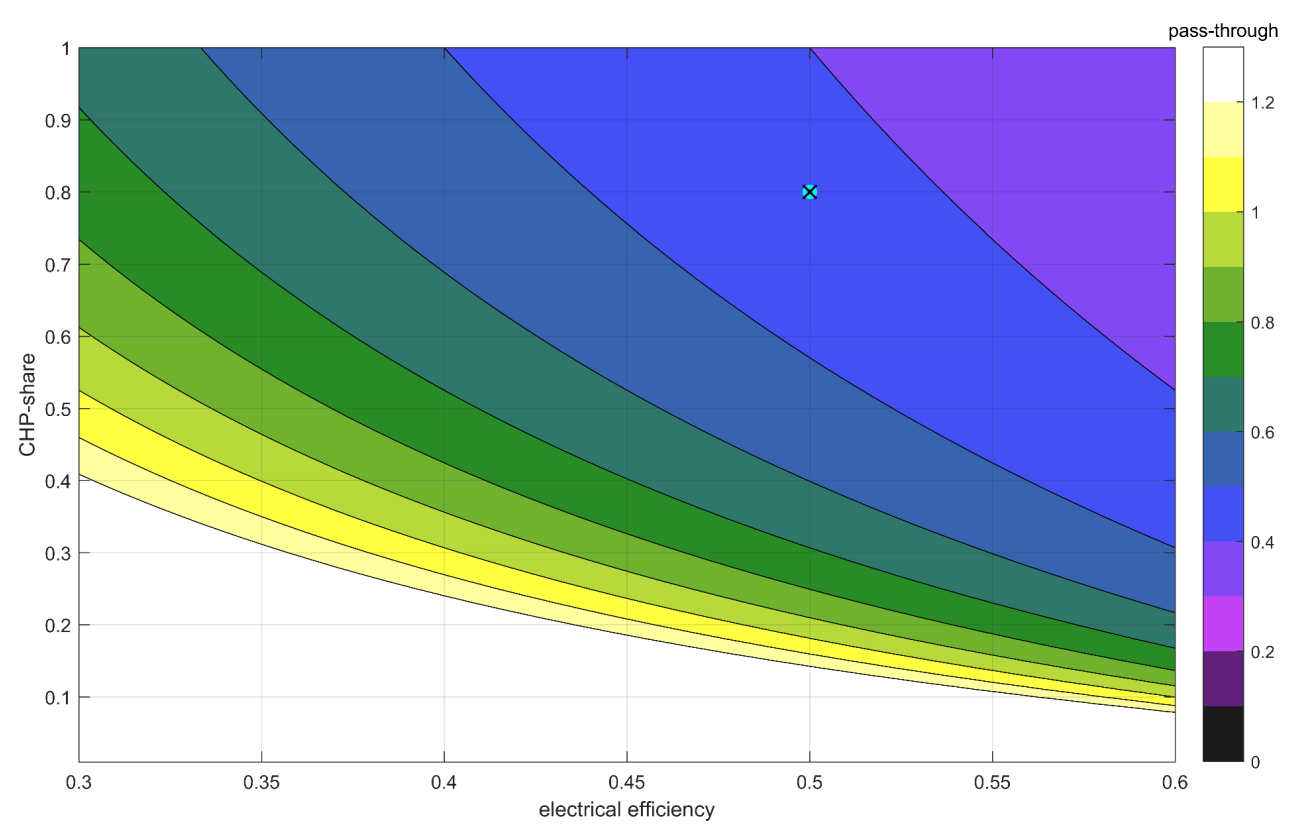}
\caption{Minimally required pass-through for total cost reduction. Dot indicates the lowest pass-through level compatible with cost reduction for a district heating system with electrical efficiency of $0.5$ and a CHP heat generation share of $0.8$.}\label{Fig1}
\end{figure}

Part $\text{A}$ of the right-hand side of (\ref{8}) is strictly positive for district heating portfolios, while the sign of $\text{B}$ depends on the difference in heat cost between boilers and CHP units.
$\text{B}$ will be zero either if CHP units are capacity constrained (i.e. $\varepsilon_{s^{C}}^{p_{el}} = 0$) or if the heat cost of boilers and CHP units is equal.
If (i) CHP units have a cost advantage (cf. equation (\ref{5})) the whole of $\text{B}$ will be negative, lowering the threshold for the minimally required pass-through from \ce{CO2} emission cost to electricity prices.
On the other hand, if (ii) electricity prices are so low that boilers have a cost advantage, heat will be preferentially generated in boilers, up to the available boiler capacity and heat generation from CHP units is at the minimal level.
In consequence, heat generation does not change with the electricity price, i.e. $\varepsilon_{s^C}^{p_{el}} = 0$.
This holds true until the electricity price reaches a level at which $k_{th}^{B} = k_{th}^{C}$ and we are back in case~(i).
Hence, we can restrict further analysis to the most adverse case (highest level of pass-through required for total cost reduction) in which B equals zero and (\ref{8}) simplifies to
\begin{align}\label{9}
\frac{\partial p_{el}}{\partial p_{e}} > \left( \frac{q^{T} - q^{C}}{q^{C}} \frac{\eta_{th}^{C}}{\eta_{th}^{B} \eta_{el}^{C}} + \frac{\eta_{th}^{B}}{\eta_{th}^{B} \eta_{el}^{C}} \right) em_{f}
\end{align}

Figure~\ref{Fig1} depicts the minimal total cost reducing pass-through for given shares of CHP heat generation in total emission intense heat generation and for given electrical efficiency of CHP units. 
As is to be expected, a high CHP share and high electrical efficiency both lead to lower required pass-through levels. 

As the required pass-through depends on technical characteristics of district heating systems, Table~\ref{Tab1} presents exemplary requirements on the minimal pass-through that guarantees total cost reduction.
In addition, Table~\ref{Tab1} also indicates the minimal electricity mark-up over fuel and emission cost that allows CHP plants to enjoy a cost advantage over heat boilers.
For district heating systems with the least efficient power generators, the minimal pass-through compatible with reduction of total cost is more than twice as high ($0.759$) as for district heating systems with the most efficient power generators ($0.352$).
More typical district heating systems (electrical efficiency $0.4$) are able to reduce their cost with rising emission prices, provided that at least 55.6\% of the increase in \ce{CO2} emission prices is passed on to electricity prices.

In the following, we use the power system model \emph{MEDEA\_{lin}} to investigate the effect of emission prices on electricity prices, i.e. to estimate the pass-through $\partial p_{el} / \partial p_{e}$.
Power plant dispatch and corresponding power prices are derived for \ce{CO2} prices of $5.89$ \euro{}/t\ce{CO2} (which is equal to the average EUA price in 2017) and all $5$ \euro{}/t\ce{CO2} increments up to $80.89$ \euro{}/t\ce{CO2}, given the current inventory of power plants.
The results are used to approximate $\partial p_{el} / \partial p_{e} \approxeq \Delta p_{el} / \Delta p_{e}$, the pass-through from emission prices to electricity prices.

\begin{table}[t]
\caption{Effects of technical parameters on heat generation units}
\smallskip
\centering
\begin{tabular}{c c c}
\hline
\makecell{Electrical \\efficiency} & \makecell{Min. mark-up \\(cf. eq. (\ref{5}))} &
\makecell{Min. pass-through \\ (cf. eq. (\ref{9}))} \\ [1.5 ex]
\hline  \hline
$0.3$ & $0.481$ & $0.759$ \\ [0.5 ex] \hline
$0.4$ & $0.389$ & $0.556$ \\ [0.5 ex] \hline
$0.5$ & $0.333$ & $0.433$ \\ [0.5 ex] \hline
$0.6$ & $0.296$ & $0.352$ \\ [1.0 ex]
\hline \hline
\end{tabular}
\label{Tab1}
\end{table}

\section{Data and Methods}
\subsection{Description of the power system Model \emph{MEDEA\_lin}}
We use a simple, stylized and parsimonious model of the Austrian and German power system to estimate the pass-through from emission prices to electricity prices in the Austrian and German electricity market.
Our power system model determines the cost-minimizing hourly dispatch of thermal and hydro storage power plants that is required to meet price-inelastic (residual) demand for electricity and district heat.
In total, $552$ thermal power plants and $62$ hydro storage plants are grouped in $34$ technology clusters which are differentiated by generation technologies (e.g. steam turbine, combustion turbine, combined cycle, etc.) and by fuels (uranium, lignite, hard coal, natural gas, mineral oil, biomass, water).
Thermal power plants must burn fuels to generate electricity and are constrained in their operation by installed capacities for heat and power generation.
Heat generation is possible in heat boilers (aggregate capacity $30$ GWth, efficiency $0.9$) or in combined heat and power (CHP) plants, which must respect the limits of their feasible operation region.\footnote{The feasible operation region specifies all viable combinations of heat and power generation along with the required fuel use.}
Heat generation and consumption are considered in aggregate, i.e. heat generators are not serving specific district heating systems.
Electricity generation from non-dispatchable renewable sources (wind, solar, run-of-river) is taken as given, but can be curtailed. 
Electricity can also be stored in reservoir and pumped hydro storages.
Generation from hydro storage plants is constrained by turbine capacity and energy contained in reservoirs of limited size.
Reservoirs are filled by inflows or by pumping (pumped storages only).
To better capture operational differences, we model daily, weekly and seasonal reservoir and pumped storage plants separately.
Electricity exchange with countries outside Austria and Germany is held fixed at hourly quantities realized in 2017.
To ensure a stable and secure operation of the electricity system, power plants must provide ancillary services (e.g. frequency control, voltage support).
We assume that this requires generators with an installed capacity of at least $21$ GW\footnote{This is equal to 12.5\% of peak load plus 7.5\% of installed solar and wind power capacity and is broadly in line with findings by \citet{Hirth2015} and \citet{Nicolosi2012}.} to be operational (either generating or pumping in case of pumped hydro storages) at any point in time.
For a mathematical description see \ref{Apndx2}. The model is implemented in GAMS and was solved by Gurobi on an Intel Xeon Gold $6144$ with $264$ GB RAM.
The model code can be found at \url{https://github.com/sebwehrle/medea}.

\subsection{Data}
Information regarding the power plant stock (generation capacities, technology, locations) in Germany are based on data provided by the Open Power System Data (OPSD) project \citep{OPSD2018a}.
Data on Austrian power plants was collected through own research and includes information from the regulatory body~\citep{EControl2003}, sector associations~\citep{OeEn2017}, operating companies, and water registers of federal authorities~\citep{LandVorarlberg2018, LandTirol2018}.
Hourly electricity generation from intermittent sources (solar PV, wind) and load in Austria and Germany is also sourced from \citet{OPSD2018b}. 
Further time series on international commercial electricity exchanges, the aggregate filling rate of hydro reservoirs and storage plants, and the actual electricity generation and consumption of hydro power plants (including run-of-river plants) are obtained from ENTSO-E's transparency platform~\citep{ENTSOE2018}.
We approximate inflows of water to reservoirs in Austria by combining downsampled data on weekly water reservoir levels with hourly electricity generation and pumping from hydro reservoirs and pumped storage plants.\footnote{In Germany, 98\% of the installed hydro storage capacity is pumped hydro power so that we abstract from inflows to German hydro storage plants.}
Hourly district heating demand is estimated based on synthetic load profiles for natural gas demand~\citep{Almbauer2009}.
These load profile make use of daily average temperatures from MERRA-2 satellite data~\citep{GMAO2015} and are scaled to total final consumption of district heat in Germany and Austria.\footnote{As final data for the consumption of district heat in 2017 was not published at the time of writing, we scale 2016 district heat consumption by the relative change in heating degree days from 2016 to 2017~\citep{AGEB2017, OeStat2018}.}
Realized prices of hard coal, natural gas and EU emission allowances for the year 2017 are taken from the \citet{eex2018}.
Prices for mineral oil are approximated on the basis of prices for Brent crude oil as published by the~\citet{EIA2018}.
As there are no market prices for nuclear fuel and lignite, we estimate lignite cost at $5.50$~\euro{}/MWh (including mining, but excluding emission cost) and $3.50$~\euro{}/MWh for nuclear fuel.
Descriptive Statistics for all used time series are displayed in Table~\ref{Tab2}.

\begin{table}[t!]
\caption{Descriptive statistics}
\smallskip
\centering
\begin{tabular}{c c c c c c c}
\hline
 & Unit & Min & Max & Mean & Std. Dev & Source \\ \hline \hline
Wind & $GW$ & $0.279$ & $41.72$ & $12.49$ & $9.03$ & OPSD \\ \hline
Solar PV & $GW$ & $0$ & $28.33$ & $4.23$ & $6.49$ & OPSD \\ \hline
Run-of-river & $GW$ & $2.08$ & $7.13$ & $4.78$ & $1.13$ & ENTSO-E \\ \hline
\makecell{Reservoir \\ inflows} & $GW$ & $0$ & $2.14$ & $0.79$ & $0.59$ & \makecell{own\\ calculations} \\ \hline
\makecell{Electricity \\consumption} & $GW$ & $38.52$ & $96.77$ & $69.09$ & $12.59$ & ENTSO-E \\ \hline
\makecell{District heat \\ consumption} & $GW$ & $1.54$ & $42.04$ & $13.49$ & $9.60$ & \makecell{own \\calculations} \\ \hline
\makecell{Net imports \\ (commercial \\exchange)} & $GW$ & $-13.74$ & $7.07$ & $-4.87$ & $2.93$ & ENTSO-E\\ \hline
\makecell{Coal price \\(API2)} & \euro{}$/MWh_{th}$ & $8.32$ & $11.36$ & $9.65$ & $0.72$ & eex \\ \hline
\makecell{Natural gas \\price (NCG)} & \euro{}$/MWh_{th}$ & $15.91$ & $19.02$ & $17.27$ & $0.87$ & eex\\ \hline
\makecell{Mineral oil \\price} & \euro{}$/MWh_{th}$ & $23.18$ & $32.93$ & $28.25$ & $2.57$ & EIA\\ \hline
\makecell{EU Emission \\Allowance price} & \euro{}$/MWh_{th}$ & $4.41$ & $8.21$ & $5.89$ & $1.11$ & eex\\
\hline \hline
\end{tabular}
\label{Tab2}
\end{table}

\subsection{Model calibration and goodness-of-fit}

\begin{figure}[t]
\centering
\includegraphics[scale=0.75]{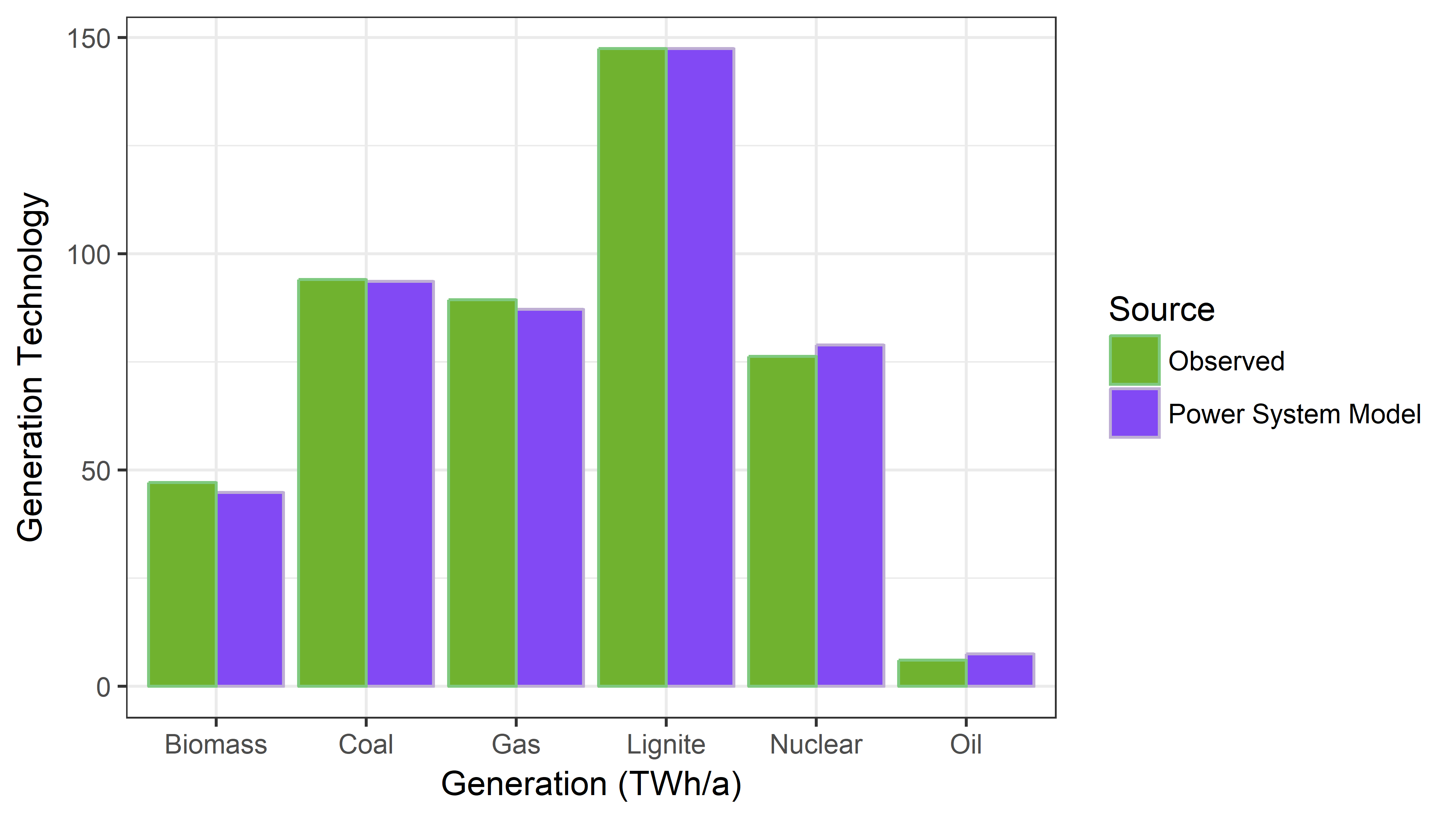}
\caption{Actual and model derived fuel burn (2017)}\label{Fig2}
\end{figure}

We calibrate our power system model with actual data on quantities of fuel burnt for power generation~\citep{AGEB2018, OeStat2018} and greenhouse gas emissions \citep{UBA_DE2018a, UBA_DE2018b} from 2017.
\footnote{For Austria, the most recent data on GHG emissions and fuel burn for power generation available at the time of writing relates to 2016. As Austria’s electricity consumption is only around one tenth of Germany’s and roughly two-third of power generation in Austria stems from hydro power plants, the induced error should, however, be small.}

As visible in figure \ref{Fig2}, our model is able to replicate fuel burn and emissions fairly well, although the use of mineral oil and nuclear fuel is somewhat overstated ($+1.3$ TWh and $+2.6$ TWh, respectively), while consumption of natural gas and biomass are underestimated by $3.4$ TWh and $2.3$ TWh. Overall, modelled \ce{CO2} emissions from power generation of $318.8$ million tonnes are falling short of near time estimates by around $1.7\%$ or $5.5$ million tonnes.
Comparing the model-derived hourly shadow prices of electricity to actual day-ahead prices at the European Energy Exchange in 2017, we observe a correlation of $0.80$ and a root mean squared error (RMSE) of $10.91$.

\section{Results}
To approximate pass-through rates from \ce{CO2} emission costs to electricity prices, we determine power plant dispatch in $16$ emission price scenarios $s = 0,1,\ldots,15$.
We start from actually observed daily EU emission allowance prices (annual average in 2017: $5.89$ \euro{}/t\ce{CO2}) and increase the price by $5$ \euro{}/t\ce{CO2} in each subsequent scenario up to an annual average price of $80.89$ \euro{}/t\ce{CO2}, leaving everything else unchanged.
The marginal on the electricity supply and demand balance equation (see equation (\ref{12}) in~\ref{Apndx2}) of our power system model reflect the endogenously determined hourly electricity prices.
We use the electricity base price $p_{el}$ (i.e. the annual average of the hourly spot price) to approximate the pass-through from emission costs to electricity prices by
\begin{align}\label{10}
\frac{\Delta p_{el}}{\Delta p_{e}} = \frac{p_{el}^{s} - p_{el}^{s-1}}{p_{e}^{s} - p_{e}^{s-1}}
\end{align}
The resulting pass-through estimates are presented in Figure \ref{Fig3}.

\subsection{System-wide effects of increasing \ce{CO2} emission prices}
\begin{figure}[t]
\centering
\includegraphics[scale=0.75]{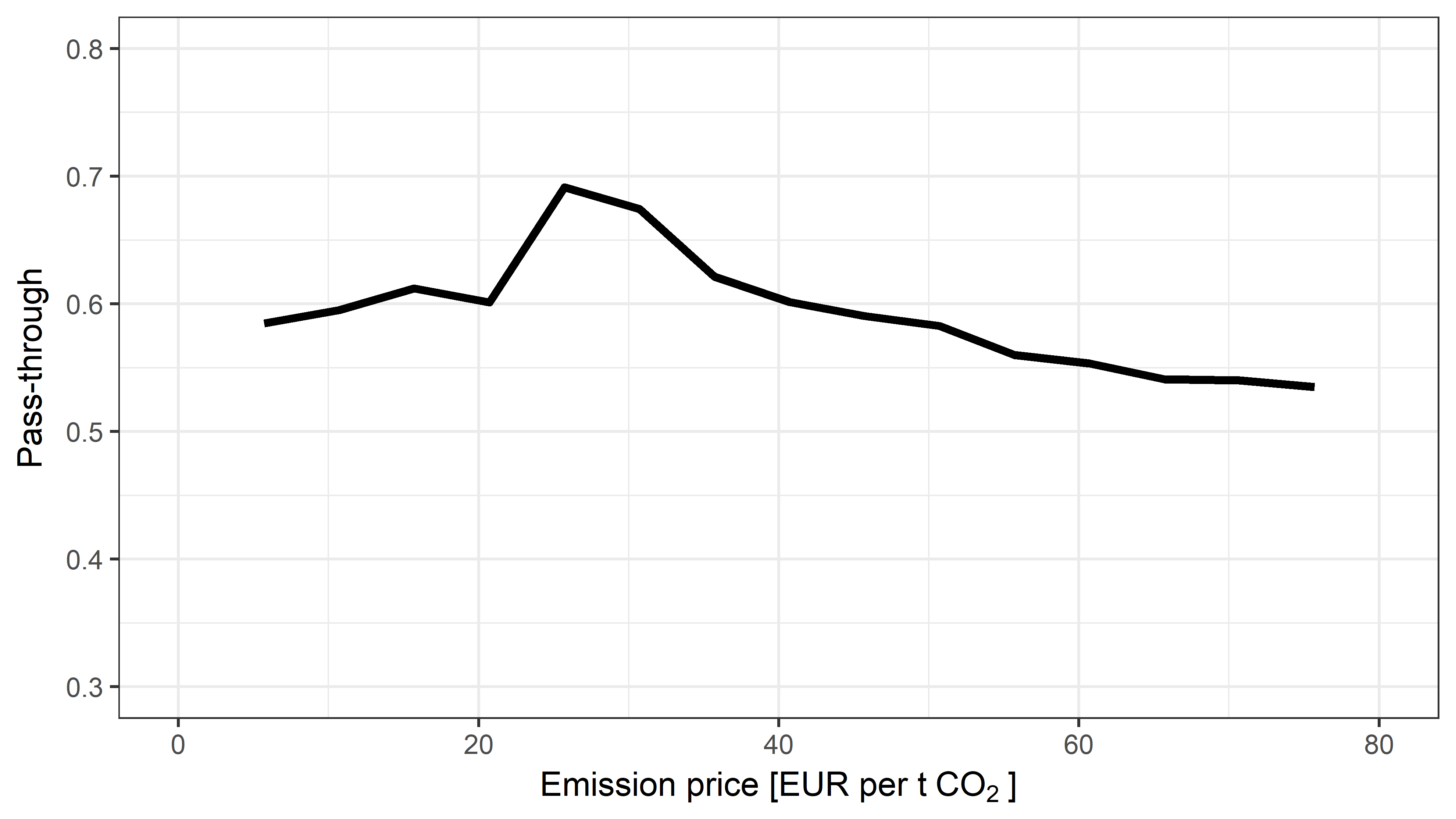}
\caption{Estimated pass-through from emission prices to electricity prices (2017)}\label{Fig3}
\end{figure}
As illustrated by Figure \ref{Fig3}, the cost pass-through depends on the absolute level of the \ce{CO2} emission price.
We estimate a pass-through close to $0.57$ at emission prices around $5$ to $10$ \euro{}/t\ce{CO2}, i.e. around $57\%$ of an increase in emission prices should be passed on to the electricity price.
With emission prices in the range of $25$-$30$~\euro{}/t\ce{CO2}, our pass-through estimate reaches its maximum of $0.68$ and gradually declines from there on.
At emission price levels close to $80$~\euro{}/t\ce{CO2} it reaches $0.53$.

The co-movement of \ce{CO2} emission prices and the pass-through can be explained by (i) a 'fuel switch' triggered by rising emissions prices, (ii) a 'CHP generation switch' and (iii) an increase in the overall fuel efficiency.

\begin{figure}[t]
\centering
\begin{subfigure}[t]{0.475\textwidth}
\centering
\includegraphics[scale=0.75]{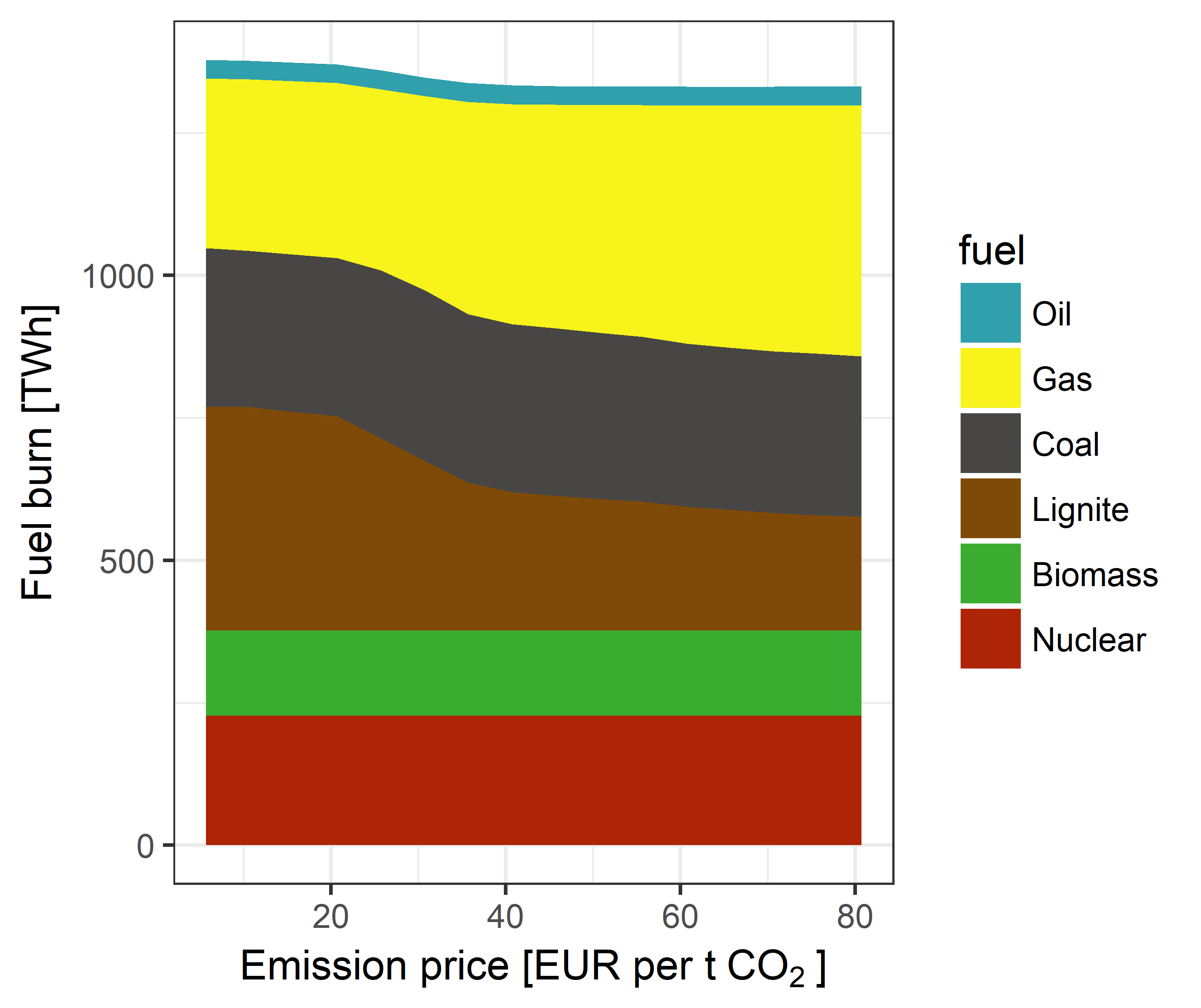}
\caption{Fuel burn for power generation} \label{Fig4a}
\end{subfigure}
\begin{subfigure}[t]{0.475\textwidth}
\centering
\includegraphics[scale=0.75]{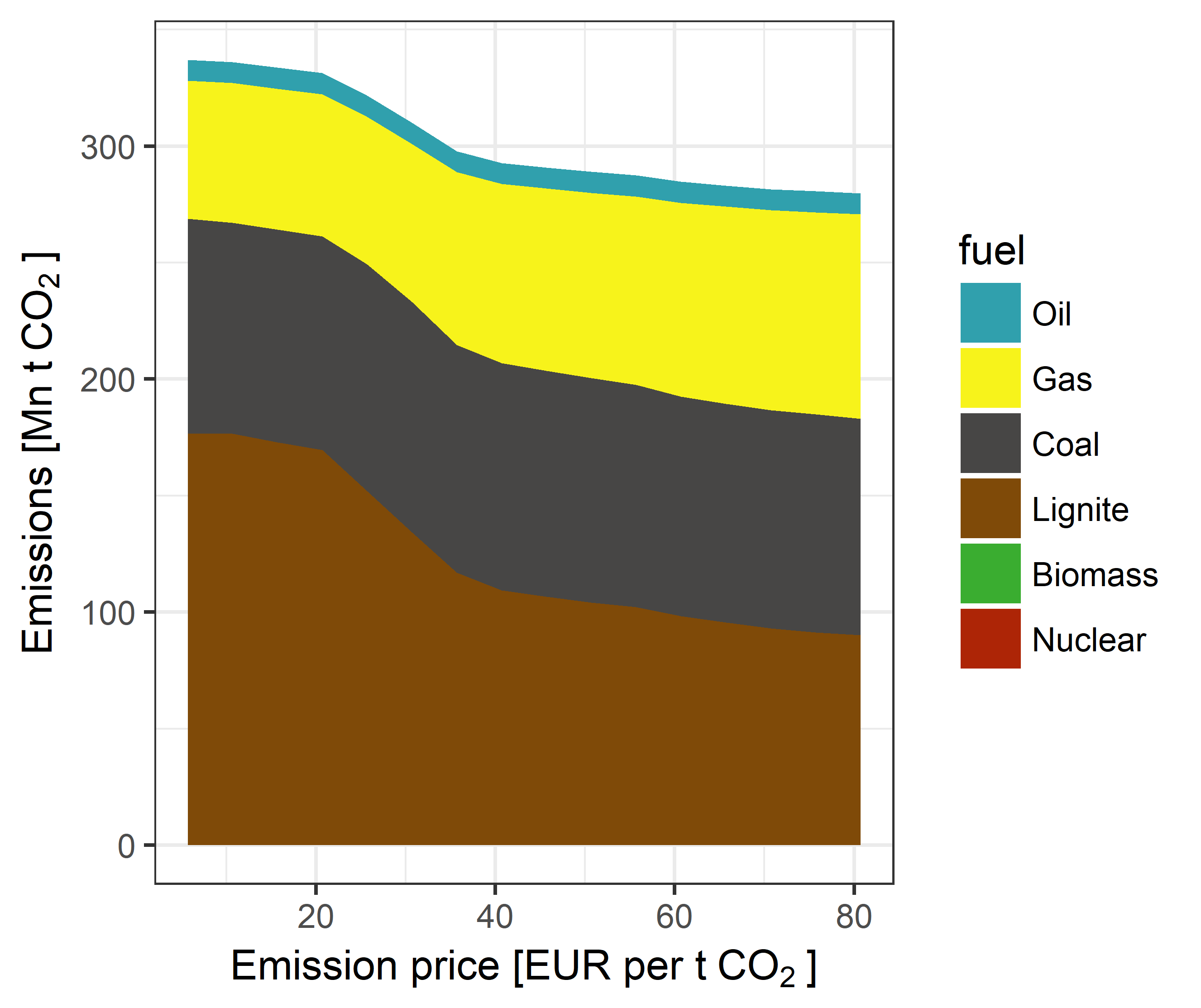}
\caption{\ce{CO2} emissions from power generation}\label{Fig4b}
\end{subfigure}
\caption{Results from emission price scenarios}\label{Fig4}
\end{figure}

The 'fuel switch' (i) in response to increasing emission costs is illustrated in Figure~\ref{Fig4a}.
At low \ce{CO2} prices, lignite and coal fired power plants dominate the generation mix, with approximately $394$ TWh of lignite and $278$ TWh of hard coal being burned over the course of the year.
Price-setting marginal power plants are predominantly fueled by coal and natural gas.
With emission prices rising towards $20$~\euro{}/t\ce{CO2}, generation cost of lignite-fired plants increase disproportionately strong, as lignite is the fuel with the highest specific \ce{CO2} emissions.
In effect, lignite-fired generation becomes more expensive relative to generation from coal and natural gas.
While some lignite-fired units are driven out of the market, the remaining units are increasingly often becoming the price setting, marginal (i.e. most expensive still dispatched) units.

At emission prices above $20$~\euro{}/t\ce{CO2}, lignite loses most of its cost advantage relative to coal and natural gas and the use of lignite for power generation begins to decline markedly.
Under these conditions, lignite-fired units are frequently the most expensive dispatched units.
In consequence, any further increase in \ce{CO2} emission prices leads to a particularly strong effect on electricity prices.
However, this effect vanishes as more and more lignite-fired generators become uncompetitive and cease generation.
As \ce{CO2} emission prices reach $40$~\euro{}/t\ce{CO2}, the decline in lignite use slows down.
Further emission reductions are caused by the substitution of lignite and coal fired plants with natural gas fired generators.
However, even for emission prices above $70$~\euro{}/t\ce{CO2} coal and lignite account for roughly two fifth of total fuel combustion for power generation as the power plant stock is fixed in our analysis and cannot adjust to higher \ce{CO2} emission costs.

The 'CHP generation switch' (ii) occurs in parallel to the fuel switch.
At low \ce{CO2} emission cost, heat is mostly generated in natural gas and, to a lesser extent, coal fired units, with a relatively small contribution from lignite fuelled CHP plants.
As emission cost rise, heat generated in natural gas-fired CHP units is increasingly substituted by heat from lignite-fueled plants.
Due to the switch from electricity generation to heat generation, the total efficiency of lignite-fired power stations increases, effectively dampening the decline in total generation ($-15\%$ as emission prices rise from $5.89$~\euro{}/t\ce{CO2} to $30.89$~\euro{}/t\ce{CO2}) that comes in line with a considerable reduction in lignite burn ($-24\%$ over the same price range).
Conversely, natural gas-fired CHP units increase their power generation at the expense of heat generation. In spite of the shift in generation, CHP units maintain relatively high total efficiency, as the $10\%$ increase in total generation requires $11\%$ more natural gas being burnt for energy generation.

This contributes to (iii) in an increase in overall fuel efficiency, as visible in Figure~\ref{Fig4a}.
Total fuel combustion for power generation declines from $1285$ TWh at low emission prices to $1238$ TWh at the highest \ce{CO2} emission prices that we investigated.
Taken together, these effects induce relatively small marginal \ce{CO2} emission reductions as long as emission prices remain below $20$~\euro{}/t\ce{CO2} (see Figure~\ref{Fig4b}).
Above this price level marginal \ce{CO2} emission reductions increase notably as generation from natural gas-fired power plants becomes increasingly cost-competitive and substitutes for lignite-fired electricity generation.
At emission prices above $40$~\euro{}/t\ce{CO2} most of the (short-run) substitution potential is exploited and marginal \ce{CO2}~emission reductions begin to recede.

\subsection{Robustness of estimates}\label{robustn}
\begin{figure}[t]
\centering
\includegraphics[scale=0.75]{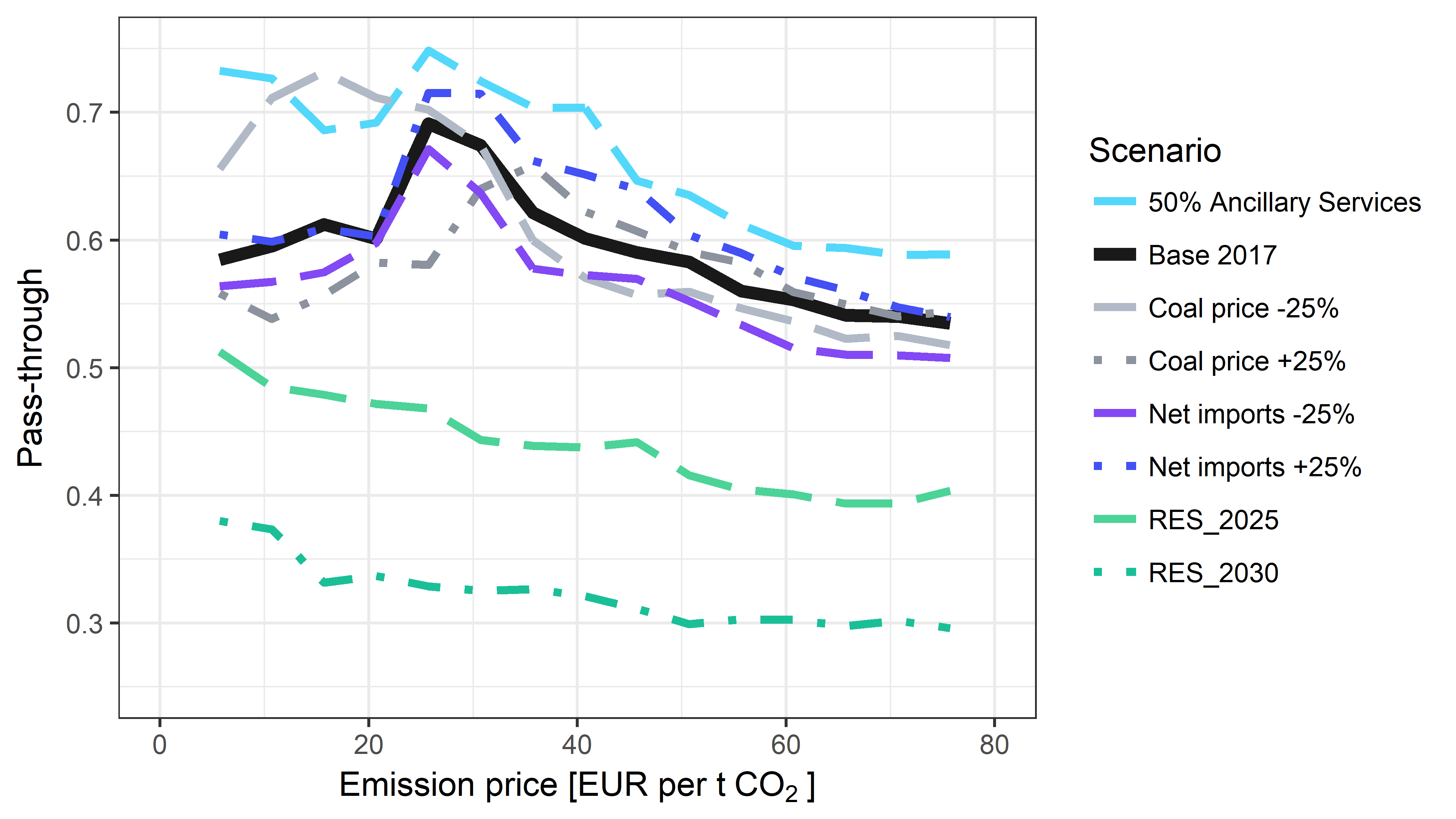}
\caption{Sensitivity of pass-through estimates}\label{Fig5}
\end{figure}

To assess the robustness of our pass-through estimates with respect to model assumptions, we conduct sensitivity runs, in which we vary the parameter(s) of interest, holding everything else constant.
Figure~\ref{Fig5} summarizes all resulting estimates for the pass-through from \ce{CO2} emission costs to electricity prices.
\subsubsection*{International electricity exchange}
An increase of $25\%$ in the (net) quantities exported and imported leads to an average increase in the emission cost pass-through by $0.02$ compared to our baseline estimate.

Germany and Austria were net exporters of electricity in 2017, i.e. over the course of the year electricity generation was higher than it would have been in the absence of international electricity exchange. 
Ceteris paribus, this marginal generation has to be sourced from thermal generators. In effect, the emission cost pass-through increases.

\subsubsection*{Relative price of coal}
We find that a $25\%$ decline in (relative) coal prices increases the emission cost pass-through by about $0.1$ at \ce{CO2} emission prices up to around $20$~\euro{}/t\ce{CO2}, as cheaper coal induces increased substitution of lignite-fired generation with coal-fired generation (instead of natural-gas fired units).
In effect, coal-fired units are more often the price-setting, marginal units and the pass-through is higher.

As the specific \ce{CO2} emissions of coal are about two-thirds higher than those of natural gas, the emission cost disadvantage of coal (versus natural gas) begins to outweighs coal's lower fuel cost at \ce{CO2} prices above $20$~\euro{}/t\ce{CO2}.
Natural gas-fired units are the dominant marginal power plants and the pass-through declines to levels in line with our base case.

\subsubsection*{Ancillary Services and system flexibility}
Lowering the capacity required for the provision of ancillary services by $50\%$ increases our estimated pass-through by $0.15$ at low and by around $0.05$ at high \ce{CO2} emission price levels.

The minimum capacity requirement for the provision of ancillary services is effective only if the residual electricity demand falls below the minimal capacity level.
In this case, supply and demand must be balanced either through curtailment of renewable electricity generation (which would raise residual demand) or through increased pumping by pumped hydro storages (which has the additional benefit that ancillary services can also be provided through pumping).
In consequence, the marginal, price-setting unit is a renewable generator and the cost of \ce{CO2} emissions is not passed through at that time.
With less capacity required to be operational for the provision of ancillary services, such incidents occur less frequently and thermal, \ce{CO2} emission intense generators are setting the price more often. This results in a higher emission cost pass-through.

\subsubsection*{Evolution of the electricity system}
Finally, we investigate the effect of the ongoing energy transition on the emission cost pass-through to electricity prices.
In line with Germany's Renewable Energy Act~\citep{EEG2017}, we assume that by electricity generation from solar PV and wind turbines will increase by $40\%$ and $50\%$, respectively up to the year 2025. In addition, we assume that $10\%$ of coal and lignite fired generation capacities will shut down by then.
Due to the increase in \ce{CO2} emission-free generation from renewable sources and the decline in \ce{CO2} emission-intense generation capacities, the emission cost pass-through declines on average by $0.15$ compared to our baseline estimation for the year 2017.

For 2030, the Renewable Energy Act envisages a further expansion of renewable energy generation. Electricity generation from solar PV should rise by $70\%$, while wind energy generation is set to double compared to the year 2017.
Together with a $20\%$ decline in coal and lignite-fired generation capacities, this lowers our pass-through estimate by $0.27$ on average, leading to pass-through levels as low as $0.3$. 

\section{Discussion}
Compared to estimates of the \ce{CO2} emission cost pass-through to electricity prices provided in the literature (see Table \ref{pttab}), our estimates are at the lower end of the findings reported by \citet{Sijm2006} and well below the range estimated by \citet{Hintermann2016} for the German market.

Yet, even with a cost pass-through as low as our estimates suggest, DHSOs operating natural gas-fired CHP plants of at least moderate electrical efficiency ($\eta_{el}^{C} \geq 0.4$) could benefit from \ce{CO2} prices up to $60$~\euro{}/t\ce{CO2}, provided that boilers generate less than $20\%$ of total heat generated from fossil fuels.\footnote{As boilers are typically used for peak generation, this should hold for many, if not most DHC systems.} (See Table \ref{Tab1})
Moderate changes in net imports should not alter pass-through substantially, as our analysis in section \ref{robustn} suggests. 
However, as we only changed the magnitude of net imports but not the pattern of (net) imports over time, our analysis might not reveal the full impact of changes in the pan-European power plant dispatch on the emission cost pass-through caused by rising \ce{CO2} emission prices.
More frequent electricity imports from low-carbon sources (e.g. French nuclear power plants) could crowd out some of the \ce{CO2} emission intense production with high marginal emission cost.
This would reduce pass-through rates.
If, on the other hand, exports would increase in response to rising \ce{CO2} prices, pass-through rates would turn out to be higher than estimated.

In the longer run, the transition towards an increasingly renewable electricity system is likely a bigger challenge for DHSOs than rising electricity imports, as our estimates suggest a considerably higher effect on emission cost pass-through.
Even highly efficient CHP plants could lose profitability if emission prices rise above $50$~\euro{}/t\ce{CO2} in the power system we assumed to be in place in the year 2030.
The rapid expansion of renewable energy generation might also be a cause for the gap between our estimates and emission cost pass-through reported  in the literature. These estimates are based on data for the German electricity system from 2005 and 2010-2013, periods with considerably lower renewable power generation than in 2017.\footnote{In 2010, $55.8$ GW of renewable capacities installed in Germany generated $105.2$ TWh of electricity. In 2013, installed capacity increased to $82.8$ GW, while gross generation rose to $152.8$ TWh. By 2017, $111.9$ GW of installed renewable capacities generated $217.9$ TWh of electricity. \citep{AGEE2018}}

Less efficient natural gas-fired CHP units are pressured by rising emission costs already today.
Although the rising emission costs should be increasingly passed through to electricity prices as emission prices rise above current levels to around $25$~\euro{}/t\ce{CO2}, our pass-through estimates are not sufficiently high to guarantee improved profitability of less efficient district heating systems.

Cheaper coal prices might provide a shelter from rising emission prices to less efficient district heating systems. However, more effective in the medium to longer run appears to be increased flexibility of the power system.

\section{Conclusions}
We have shown that emission cost pass-through to electricity prices is an important factor in the profitability of district heating systems and that emission cost pass-through with heterogeneous suppliers, such as in the electricity system, is affected by a wide variety of factors, including the emission price level itself.
In addition, our estimates of the emission cost pass-through suggest that any increase in emission cost is currently distributed roughly even between producers and consumers of electricity. 
In the longer run, however, we expect the renewables expansion to shift the odds of rising emission prices in favour of consumers, so that emission-intense producers bear the brunt of any policy that aims to reduce GHG emissions by increasing their cost.
District heating system operators could reduce the risk they face from rising emission prices by investing in more efficient or emission free co-generation technologies. 
Such investments are easier to finance as long as rising emission costs translate into increasing profits. Hence, from a policy perspective the window of opportunity, in which many DHSOs have an incentive to support measures leading to higher \ce{CO2} emission prices is closing at the speed of the renewables expansion. 
Increasing power system flexibility - nowadays a popular demand from scientists and politicians alike - could enlarge this window of opportunity. However, this comes at the cost of increasing the emission cost burden on consumers. 

\section*{Acknowledgements}
We gratefully acknowledge support from the European Research Council (“reFUEL" ERC-2017-STG 758149).

\newpage
\bibliography{passtru}

\newpage
\appendix
\section{Description of the power system model \emph{MEDEA\_lin}}\label{Apndx2}
\subsection{List of Sets}
\begin{tabular}{c l}
$f \in F$ & fuels \\
$g \in G$ & generators \\
$CHP \in G$ & combined heat and power generators \\
$PSP \in G$ & hydro storage generators \\
$l \in L$ & limits to the feasible operation region of CHP plants \\
$p \in P$ & products generated \\
$t \in T$ & time periods (hours)
\end{tabular}

\subsection{List of Parameters}
\begin{tabular}{c l}
$\underline{a}$ & \makecell[l]{minimum generation level required for provision of \\ancillary services [MW]} \\
$\overline{C}_{g,p}$ & \makecell[l]{maximal generation of product $p$ by generator in \\cluster $g$ [MW]} \\
$\overline{STOR}_{g}$ & \makecell[l]{reservor storage capacity of hydro storage plant $g$ [MWh]} \\
$D_{t,p}$ & demand for product $p$ at time $t$ [MW] \\
$N_{g}$ & number of plants in cluster $g$ \\
$om_g$  & \makecell[l]{variable operation and maintenance cost of \\generator $g$ [\euro/MWh]} \\
$ORF_{g,l,f}$ & \makecell[l]{use of fuel $f$ at limit $l$ of the operating region of a \\CHP unit in cluster $g$ [MWh]} \\
$ORP_{g,l,p}$ & \makecell[l]{generation of product $p$ at limit $l$ of the operating \\region of a CHP unit in cluster $g$ [MWh]} \\
$P_{t,eua}$ & price of \ce{CO2} emissions at time $t$ [\euro/t~\ce{CO2}e] \\
$P_{t,f}$ & price of fuel $f$ at time $t$ [\euro/MWh] \\
$Q_{t,pv}$ & solar energy generated at time $t$ [MW] \\
$Q_{t,ror}$ & energy generated by run-of-river plants at time $t$ [MW]\\
$Q_{t,we}$ & wind energy generated at time $t$ [MW]\\
$Q_{t,nip}$ & net imports at time $t$ [MWh] \\
$Q_{t,res}$ & inflows into hydro reservoirs at time $t$ [MWh/h]\\
$em_{f}$ & emission factor of fuel $f$ [t \ce{CO2}/MWh fuel used]\\
$\eta_{g,f,p}$ & \makecell[l]{efficiency of generator $g$ using fuel $f$ to generate product $p$}\\

\end{tabular}

\subsection{List of Variables}
\begin{tabular}{c l}
$ppsp_{t,g}$ & \makecell[l]{energy pumped into pumped storage reservoirs at \\time $t$ [MWh]} \\
$qf_{t,g,f}$ & quantity of fuel $f$ used by generator $g$ at time $t$ [MWh]\\
$qp_{t,g,p}$ & energy generated by generator $g$ at time $t$ [MWh]\\
$qpsp_{t,g}$ & power generated by hydro storage plant $g$ at time $t$ [MWh]\\
$sconv_{t,g,l}$ & convexity variable $g$\\
$qstor_{t,g}$ & \makecell[l]{quantity of energy stored in hydro storage plant $g$ at \\time $t$ [MWh]}\\
$qns_{t,p}$ & non-served load of product $p$ at time $t$ [MWh]\\
$qct_{t}$ & quantity of energy curtailed at time $t$ [MWh]\\
\end{tabular}

\subsection{Mathematical description of the power system model MEDEA\_lin}
Our power system model uses a linear programming formulation of the economic dispatch problem for thermal units within the Austro-German bidding zone. Operation of pumped storage plants is also formulated as a linear problem.
The model's objective is to minimize total system cost, the sum of fuel, emission and operation and maintenance cost along with the cost associated with curtailment of renewable energies and loss of load.
\begin{align}
\min \left(\sum_{t,g,f}\left(\left(P_{t,f} + em_{f} P_{t,eua}+om_{g}\right) qf_{t,g,f} + qns_{t,p} M + qct_{t} N \right) \right)
\end{align}
In each hour the market has to clear, such that electricity supply from thermal and net generation from hydro storage plants plus power generation from non-dispatchable sources (wind energy, photovoltaics, and run-of-river hydro plants) equals electricity demand less net imports of electricity.
\begin{align}\label{12}
\begin{split}
D_{t,pwr} - Q_{t,nip} = \sum_{g} \left(q p_{t,g,pwr} \right) + \sum_{g\in PSP} \left(qpsp_{t,g} - ppsp_{t,g} \right) \\
+ Q_{t,we} + Q_{t,pv} + Q_{t,ror}, \forall t
\end{split}
\end{align}
In linear (economic dispatch) models, the marginals ('shadow prices') on equation (\ref{12}) can be interpreted as power prices in an energy-only market. We use these marginals to derive the pass-through from emission prices to power prices.
As we also consider co-generation of heat and power in our model, we introduce the heat balance equation (\ref{13}). Heat supply from CHP units and heat boilers must be adequate to meet district heating demand $D_{t,ht}$.
\begin{align}\label{13}
D_{t,ht} \leq \sum_{g \in CHP} \left( qp_{t,g,ht} \right) + qns_{t,p}, \forall t
\end{align}
Hourly generation of power and heat is constrained by installed capacity. 
\begin{align}
qp_{t,g,p} \leq \bar{C}_{g,p}, \forall t
\end{align}
Coproduction of heat and power in CHP-plants is governed by
\begin{align}
\sum_{l} sconv_{t,g,l} = 1, \forall g \in CHP \\
\sum_{l} sconv_{t,g,l} ORP_{g,l,p} = qp_{t,g,p}, \forall g \in CHP \\
\sum_{l} sconv_{t,g,l} ORF_{g,l,f} \leq qfuel_{t,g,f}
\end{align}
Power production by power-only generators is modelled as a linear function of plant efficiency
\begin{align}
qp_{t,g,pwr} \leq \sum_{f} \eta_{g,f,pwr} qf_{t,g,f}, \forall t, g \notin CHP  
\end{align}
Ancillary services must be provided by operating thermal generation units or active hydro storage plants (regardless of wether they are pumping or turbining).
\begin{align}
\underline{a} \leq qp_{t,g,pwr} + qpsp_{h,g} + qstor_{t,g}, \forall t
\end{align}
Operation of pumped storage plants is subject to the equations
\begin{align}
qpsp_{h,g} \eta_{g} \leq \overline{C}_{g}, \forall t, g \in PSP \\
ppsp_{t,g} \leq \overline{C}_{g}, \forall t, g \in PSP \\
qstor_{t,g} - qstor_{t-1, g} = ppsp_{h,g} \eta_{g} - qpsp_{h,g} \\
qstor_{t,g} \leq \overline{STOR}_{g}, \forall t, g \in PSP
\end{align}

\end{document}